\documentclass[superscriptaddress,aps,prc,showkeys,showpacs,twocolumn,10pt]{revtex4-1}
\usepackage[OT1,T1]{fontenc}
\usepackage{graphicx}
\usepackage{rotating}
\begin{document}

\title{An Isotensor Dibaryon in the $pp \to pp\pi^+\pi^-$ Reaction?}  
\date{\today}


\newcommand*{\IKPUU}{Division of Nuclear Physics, Department of Physics and 
 Astronomy, Uppsala University, Box 516, 75120 Uppsala, Sweden}
\newcommand*{\ASWarsN}{Nuclear Physics Division, National Centre for 
 Nuclear Research, ul.\ Hoza~69, 00-681, Warsaw, Poland}
\newcommand*{\IPJ}{Institute of Physics, Jagiellonian University, prof.\ 
 Stanis{\l}awa {\L}ojasiewicza~11, 30-348 Krak\'{o}w, Poland}
\newcommand*{\Edinb}{School of Physics and Astronomy, The University of
  Edinburgh, James Clerk Maxwell Building, Peter Guthrie Tait Road, Edinburgh
  EH9 3FD, Great Britain}
\newcommand*{\MS}{Institut f\"ur Kernphysik, Westf\"alische 
 Wilhelms--Universit\"at M\"unster, Wilhelm--Klemm--Str.~9, 48149 M\"unster, 
 Germany}
\newcommand*{\ASWarsH}{High Energy Physics Division, National Centre for 
 Nuclear Research, ul.\ Hoza~69, 00-681, Warsaw, Poland}
\newcommand*{\Budker}{Budker Institute of Nuclear Physics of SB RAS, 
 11~Acad.\ Lavrentieva Pr., Novosibirsk, 630090 Russia}
\newcommand*{\Novosib}{Novosibirsk State University, 2~Pirogova Str., 
 Novosibirsk, 630090 Russia}
\newcommand*{\PGI}{Peter Gr\"unberg Institut, PGI--6 Elektronische 
 Eigenschaften, Forschungszentrum J\"ulich, 52425 J\"ulich, Germany}
\newcommand*{\DUS}{Institut f\"ur Laser-- und Plasmaphysik, Heinrich Heine 
 Universit\"at D\"usseldorf, Universit\"atsstr.~1, 40225 D\"usseldorf, Germany}
\newcommand*{\IFJ}{The Henryk Niewodnicza{\'n}ski Institute of Nuclear 
 Physics, Polish Academy of Sciences, ul.\ Radzikowskiego~152, 31-342 
 Krak\'{o}w, Poland}
\newcommand*{\PITue}{Physikalisches Institut, Eberhard Karls Universit\"at 
 T\"ubingen, Auf der Morgenstelle~14, 72076 T\"ubingen, Germany}
\newcommand*{\Kepler}{Kepler Center for Astro and Particle Physics,
  Physikalisches Institut der Universit\"at T\"ubingen, Auf der 
 Morgenstelle~14, 72076 T\"ubingen, Germany}
\newcommand*{\IKPJ}{Institut f\"ur Kernphysik, Forschungszentrum J\"ulich, 
 52425 J\"ulich, Germany}
\newcommand*{\ZELJ}{Zentralinstitut f\"ur Engineering, Elektronik und 
 Analytik, Forschungszentrum J\"ulich, 52425 J\"ulich, Germany}
\newcommand*{\Erl}{Physikalisches Institut, Friedrich--Alexander Universit\"at
 Erlangen--N\"urnberg, Erwin--Rommel-Str.~1, 91058 Erlangen, Germany}
\newcommand*{\ITEP}{Institute for Theoretical and Experimental Physics named 
 by A.I.\ Alikhanov of National Research Centre ``Kurchatov Institute'', 
 25~Bolshaya Cheremushkinskaya Str., Moscow, 117218 Russia}
\newcommand*{\Giess}{II.\ Physikalisches Institut, 
 Justus--Liebig--Universit\"at Gie{\ss}en, Heinrich--Buff--Ring~16, 35392 
 Giessen, Germany}
\newcommand*{\IITI}{Discipline of Physics, Indian Institute of Technology 
 Indore, Khandwa Road, Indore, Madhya Pradesh 453 552, India}
\newcommand*{\HepGat}{High Energy Physics Division, Petersburg Nuclear Physics 
 Institute named by B.P.\ Konstantinov of National Research Centre ``Kurchatov 
 Institute'', 1~mkr.\ Orlova roshcha, Leningradskaya Oblast, Gatchina, 188300
 Russia}
\newcommand*{\HeJINR}{Veksler and Baldin Laboratory of High Energiy Physics, 
 Joint Institute for Nuclear Physics, 6~Joliot--Curie, Dubna, 141980 Russia}
\newcommand*{\Katow}{August Che{\l}kowski Institute of Physics, University of 
  Silesia, ul.\ 75 Pu{\l}ku Piechoty 1, 41-500 Chorz\'{o}w, Poland}
\newcommand*{\NITJ}{Department of Physics, Malaviya National Institute of 
 Technology Jaipur, JLN Marg, Jaipur, Rajasthan 302 017, India}
\newcommand*{\JARA}{JARA--FAME, J\"ulich Aachen Research Alliance, 
 Forschungszentrum J\"ulich, 52425 J\"ulich, and RWTH Aachen, 52056 Aachen, 
 Germany}
\newcommand*{\Bochum}{Institut f\"ur Experimentalphysik I, Ruhr--Universit\"at 
 Bochum, Universit\"atsstr.~150, 44780 Bochum, Germany}
\newcommand*{\IITB}{Department of Physics, Indian Institute of Technology 
 Bombay, Powai, Mumbai, Maharashtra 400 076, India}
\newcommand*{\Tomsk}{Department of Physics, Tomsk State University, 36~Lenin 
 Ave., Tomsk, 634050 Russia}
\newcommand*{\KEK}{High Energy Accelerator Research Organisation KEK, Tsukuba, 
 Ibaraki 305--0801, Japan} 
\newcommand*{\ASLodz}{Astrophysics Division, National Centre for Nuclear
 Research, Box~447, 90-950 {\L}\'{o}d\'{z}, Poland}

\author{P.~Adlarson}    \affiliation{\IKPUU}
\author{W.~Augustyniak} \affiliation{\ASWarsN}
\author{W.~Bardan}      \affiliation{\IPJ}
\author{M.~Bashkanov}   \affiliation{\Edinb}
\author{F.S.~Bergmann}  \affiliation{\MS}
\author{M.~Ber{\l}owski}\affiliation{\ASWarsH}
\author{A.~Bondar}      \affiliation{\Budker}\affiliation{\Novosib}
\author{M.~B\"uscher}   \affiliation{\PGI}\affiliation{\DUS}
\author{H.~Cal\'{e}n}   \affiliation{\IKPUU}
\author{I.~Ciepa{\l}}   \affiliation{\IFJ}
\author{H.~Clement}     \affiliation{\PITue}\affiliation{\Kepler}
\author{E.~Czerwi{\'n}ski}\affiliation{\IPJ}
\author{K.~Demmich}     \affiliation{\MS}
\author{R.~Engels}      \affiliation{\IKPJ}
\author{A.~Erven}       \affiliation{\ZELJ}
\author{W.~Erven}       \affiliation{\ZELJ}
\author{W.~Eyrich}      \affiliation{\Erl}
\author{P.~Fedorets}    \affiliation{\IKPJ}\affiliation{\ITEP}
\author{K.~F\"ohl}      \affiliation{\Giess}
\author{K.~Fransson}    \affiliation{\IKPUU}
\author{F.~Goldenbaum}  \affiliation{\IKPJ}
\author{A.~Goswami}     \affiliation{\IKPJ}\affiliation{\IITI}
\author{K.~Grigoryev}   \affiliation{\IKPJ}\affiliation{\HepGat}
\author{L.~Heijkenskj\"old}\altaffiliation[present address: ]{\Mainz}\affiliation{\IKPUU}
\author{V.~Hejny}       \affiliation{\IKPJ}
\author{N.~H\"usken}    \affiliation{\MS}
\author{L.~Jarczyk}     \affiliation{\IPJ}
\author{T.~Johansson}   \affiliation{\IKPUU}
\author{B.~Kamys}       \affiliation{\IPJ}
\author{G.~Kemmerling}\altaffiliation[present address: ]{\JCNS}\affiliation{\ZELJ}
\author{A.~Khoukaz}     \affiliation{\MS}
\author{O.~Khreptak}    \affiliation{\IPJ}
\author{D.A.~Kirillov}  \affiliation{\HeJINR}
\author{S.~Kistryn}     \affiliation{\IPJ}
\author{H.~Kleines}\altaffiliation[present address: ]{\JCNS}\affiliation{\ZELJ}
\author{B.~K{\l}os}     \affiliation{\Katow}
\author{W.~Krzemie{\'n}}\affiliation{\ASWarsH}
\author{P.~Kulessa}     \affiliation{\IFJ}
\author{A.~Kup\'{s}\'{c}}\affiliation{\IKPUU}\affiliation{\ASWarsH}
\author{K.~Lalwani}     \affiliation{\NITJ}
\author{D.~Lersch}\altaffiliation[present address: ]{\FSU}\affiliation{\IKPJ}
\author{B.~Lorentz}     \affiliation{\IKPJ}
\author{A.~Magiera}     \affiliation{\IPJ}
\author{R.~Maier}       \affiliation{\IKPJ}\affiliation{\JARA}
\author{P.~Marciniewski}\affiliation{\IKPUU}
\author{B.~Maria{\'n}ski}\affiliation{\ASWarsN}
\author{H.--P.~Morsch}  \affiliation{\ASWarsN}
\author{P.~Moskal}      \affiliation{\IPJ}
\author{H.~Ohm}         \affiliation{\IKPJ}
\author{W.~Parol}       \affiliation{\IFJ}
\author{E.~Perez del Rio}\altaffiliation[present address: ]{\INFN}\affiliation{\PITue}\affiliation{\Kepler}
\author{N.M.~Piskunov}  \affiliation{\HeJINR}
\author{D.~Prasuhn}     \affiliation{\IKPJ}
\author{D.~Pszczel}     \affiliation{\IKPUU}\affiliation{\ASWarsH}
\author{K.~Pysz}        \affiliation{\IFJ}
\author{J.~Ritman}\affiliation{\IKPJ}\affiliation{\JARA}\affiliation{\Bochum}
\author{A.~Roy}         \affiliation{\IITI}
\author{Z.~Rudy}        \affiliation{\IPJ}
\author{O.~Rundel}      \affiliation{\IPJ}
\author{S.~Sawant}      \affiliation{\IITB}
\author{S.~Schadmand}   \affiliation{\IKPJ}
\author{I.~Sch\"atti--Ozerianska}\affiliation{\IPJ}
\author{T.~Sefzick}     \affiliation{\IKPJ}
\author{V.~Serdyuk}     \affiliation{\IKPJ}
\author{B.~Shwartz}     \affiliation{\Budker}\affiliation{\Novosib}
\author{T.~Skorodko}\affiliation{\PITue}\affiliation{\Kepler}\affiliation{\Tomsk}
\author{M.~Skurzok}     \affiliation{\IPJ}
\author{J.~Smyrski}     \affiliation{\IPJ}
\author{V.~Sopov}       \affiliation{\ITEP}
\author{R.~Stassen}     \affiliation{\IKPJ}
\author{J.~Stepaniak}   \affiliation{\ASWarsH}
\author{E.~Stephan}     \affiliation{\Katow}
\author{G.~Sterzenbach} \affiliation{\IKPJ}
\author{H.~Stockhorst}  \affiliation{\IKPJ}
\author{H.~Str\"oher}   \affiliation{\IKPJ}\affiliation{\JARA}
\author{A.~Szczurek}    \affiliation{\IFJ}
\author{A.~Trzci{\'n}ski}\affiliation{\ASWarsN}
\author{M.~Wolke}       \affiliation{\IKPUU}
\author{A.~Wro{\'n}ska} \affiliation{\IPJ}
\author{P.~W\"ustner}   \affiliation{\ZELJ}
\author{A.~Yamamoto}    \affiliation{\KEK}
\author{J.~Zabierowski} \affiliation{\ASLodz}
\author{M.J.~Zieli{\'n}ski}\affiliation{\IPJ}
\author{J.~Z{\l}oma{\'n}czuk}\affiliation{\IKPUU}
\author{P.~{\.Z}upra{\'n}ski}\affiliation{\ASWarsN}
\author{M.~{\.Z}urek}   \affiliation{\IKPJ}

\newcommand*{\Mainz}{Institut f\"ur Kernphysik, Johannes 
 Gutenberg Universit\"at Mainz, Johann--Joachim--Becher Weg~45, 55128 Mainz, 
 Germany}
\newcommand*{\JCNS}{J\"ulich Centre for Neutron Science JCNS, 
 Forschungszentrum J\"ulich, 52425 J\"ulich, Germany}
\newcommand*{\FSU}{Department of Physics, Florida State University,
  77~Chieftan Way, Tallahassee, FL~32306-4350, USA}
\newcommand*{\INFN}{INFN, Laboratori Nazionali di Frascati, Via E. Fermi, 40, 
 00044 Frascati (Roma), Italy}

\collaboration{WASA-at-COSY Collaboration}\noaffiliation

\begin{abstract}

Exclusive measurements of the quasi-free $pp \to pp\pi^+\pi^-$ reaction have
been carried out at WASA@COSY by means of $pd$ collisions at $T_p$ = 1.2
GeV. Total and differential cross sections have been extracted covering the
energy region $T_p = 1.08 - 1.36$ GeV, which is the region of $N^*(1440)$ and
$\Delta(1232)\Delta(1232)$ resonance excitations. Calculations describing these
excitations by $t$-channel meson exchange are at variance with the measured
differential cross sections and underpredict substantially the experimental
total cross section. An isotensor $\Delta N$ dibaryon resonance with $I(J^P) =
2(1^+)$ produced associatedly with a pion is able to overcome these
deficiencies. 

\end{abstract}

\maketitle

\section{Introduction}

  Multi-quark states like tetra-, penta- and hexaquark (dibaryon) systems, be
  it of compact or molecule-like strcuture, are a topical issue at present
  extending largely our quark-based view of hadrons \cite{BBC}. The existence
  of dibaryons has far-reaching consequences, {\it e.g.} for the formation of
  neutron stars \cite{nstar}.
Within systematic studies of two-pion production in nucleon-nucleon ($NN$)
collisions at CELSIUS \cite{WB2,JJ,JP,TS,iso,FK,deldel,nnpipi,prl2009} and COSY
\cite{tt,evd,AE,prl2011,poldpi0pi0,isofus,pp0-,np00} the first clear-cut
evidence for a dibaryon resonance with $I(J^P) = 0(3^+)$ was observed recently
in the $pn \to d\pi^0\pi^0$ reaction 
\cite{prl2009,prl2011,poldpi0pi0}. Subsequent measurements of 
all relevant two-pion production channels
\cite{isofus,pp0-,np00,Hades,Jerus,np+-} revealed that all channels, which
contain isoscalar 
contributions, exhibit a signal of this resonance --- called now $d^*(2380)$
after observation of its pole in $pn$ scattering \cite{np,npfull,RWnew}. Its
structure is presently heavily disputed in various theoretical investigations
\cite{Gal,Kukulin,Lue,Wang}. Remarkably, 
it corresponds very well to $D_{IJ} = D_{03}$ predicted already in 1964 by
Dyson and Xuong \cite{Dyson} as 
one of six non-strange dibaryon states. Other members of that dibaryon multiplet
are the deuteron groundstate ($D_{01}$), the virtual $^1S_0$ state ($D_{10}$)
as well as the $\Delta N$ threshold states $D_{12}$ and $D_{21}$ --- with the
latter one being still purely hypothetical. But also recent state-of-the-art
Faddeev calculations predict the existence of these states \cite{GG}.

According to the standard theoretical description, the two-pion production
process at the energies of interest here is dominated by $t$-channel meson
exchange leading to excitation and decay of the Roper resonance $N^*(1440)$
and of the $\Delta(1232)\Delta(1232)$ system \cite{Luis,Zou}. Whereas in the
near-threshold region the Roper process dominates, the $\Delta\Delta$ process
takes over at incident energies beyond 1~GeV. Such calculations give quite a
reasonable description of the data, if for the Roper resonance the up-to-date
decay branchings \cite{SarantsevRoper,PDG} are used and if the $\rho$ exchange
contribution of the $\Delta\Delta$ process is tuned to describe
quantitatively the $pp \to pp\pi^0\pi^0$ data ("modified Valencia"
calculations) \cite{deldel} --- and if in the
$pn$-induced channels the $d^*(2380)$ resonance is taken into account.

However, in reexamining the $pp$-induced two-pion production channels we find
that for the $pp \to pp\pi^+\pi^-$ reaction beyond 0.9 GeV the calculated
cross sections come out now much too low (see
dashed line in Fig.~\ref{fig4}). The reason are the underlying isospin
relations between the various two-pion production channels. The purely
isospin-based prediction obtained from isospin decomposition of $pp$-induced
two-pion production \cite{iso} is shown by the shaded band in Fig. 1. The
small differences between model calculation and isospin prediction are due to
the neglect of small terms in the latter. For details see Ref. \cite{D21full}. 

The discrepancy in the $pp\pi^+\pi^-$ cross section appears just
in the region, where the isotensor dibaryon state $D_{21}$
with $I(J^P) = 2(1^+)$ was predicted by Dyson and Xuong \cite{Dyson} and
recently calculated by Gal and Garcilazo \cite{GG}.

Since all $pp \to pp\pi^+\pi^-$
data beyond 0.8 GeV stem from early low-statistics bubble-chamber
measurements \cite{Dakhno,Brunt,Shimizu,Sarantsev,Eisner,Pickup,kek}, it
appeared appropriate to reinvestigate this region by exclusive and
kinematically complete measurements.

\section{Experiment}

The $pp \to pp\pi^+\pi^-$ reaction was measured by use of the quasifree
process in $pd$ collisions. The experiment was carried out at COSY
(Forschungszentrum J\"ulich) with the 
WASA detector setup by using a proton beam of lab energy 
$T_p$~=~1.2~GeV impinging on a deuterium pellet target \cite{barg,wasa}. By
exploiting the quasi-free scattering process $p d \to pp\pi^+\pi^- +
n_{spectator}$, we cover the energy region $T_p = 1.08 - 1.36$ GeV
corresponding to $\sqrt s$ = 2.35 - 2.46 GeV.

The hardware trigger utilized in this analysis required two  
charged hits in the forward detector as well as two recorded hits in the
central detector.  

The quasi-free reaction $p d \to pp \pi^+\pi^- + n_{spectator}$
was selected in the offline analysis by requiring two proton tracks in
the forward detector as  well as a $\pi^+$ and $\pi^-$ track in the central
detector.   

That way, the non-measured spectator four-momentum could be reconstructed by a
kinematic fit with one over-constraint. The achieved resolution in $\sqrt s$
was about 20 MeV.

The charged particles registered in the segmented forward detector of WASA
have been 
identified by use of the $\Delta E - E$ energy loss method. For its
application in the data analysis, all combinations of signals stemming from the
five layers of the forward range hodoscope have been used. The charged
particles in the central detector have been identified by their curved track
in the magnetic field as well as by their energy loss in the surrounding
plastic scintillator barrel and electromagnetic calorimeter.

The requirement that the two protons have to be in the angular range
covered by the forward detector and that two pions have to be within the
angular range of the central detector reduces the overall acceptance to about
30$\%$.  The total reconstruction efficiency including all cuts and
kinematical fit has been 1.1$\%$. In total a sample of about 26000 
$pp\pi^+\pi^-$ events has been selected, which satisfy all cuts and conditions.

Efficiency and acceptance corrections of the data have been performed by MC
simulations of reaction process and detector setup. For the MC simulations
pure phase-space and model descriptions have been used. The latter will be
discussed in the next section. Since WASA does not cover the full reaction
phase space, albeit a large fraction of it, these 
corrections are not fully model independent. The hatched grey histograms in 
Figs.~\ref{fig5} - \ref{fig6} give an estimate for these systematic
uncertainties. As a measure of 
these we take the difference between model corrected results and those
obtained by assuming the "modified Valencia" calculations for the acceptance

The absolute normalization
of the data has been obtained by comparison of the simultaneously measured 
quasi-free single pion production process $pd \to pp \pi^0 + n_{spectator}$
to previous bubble-chamber results for the $pp
\to pp \pi^0$ reaction \cite{Shimizu,Eisner}. That way, the uncertainty in the
absolute normalization of our data is essentially that of the previous $pp \to
pp \pi^0$ data, {\it i.e.} in the order of 5 - 15$\%$. Details of the data
analysis and of the interpretation are given in Ref. \cite{D21full}.

\section{Results and Discussion}

\begin{figure} 
\centering
\includegraphics[width=0.8\columnwidth]{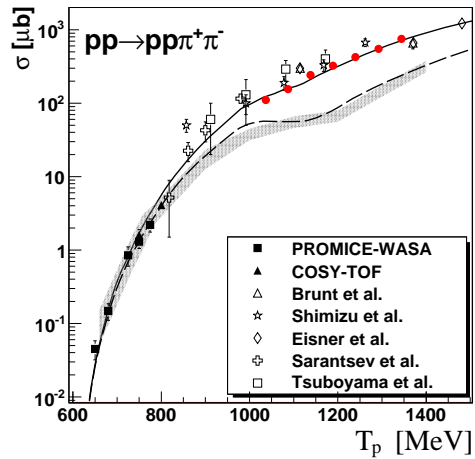}
\caption{\small (Color online) 
  Total cross section in dependence of the incident proton energy $T_p$ for
  the reaction $pp \to pp\pi^+\pi^-$. The solid dots show results from this
  work. Other symbols denote results from previous measurements
  \cite{Brunt,Shimizu,Sarantsev,Eisner,kek,WB2,JJ,JP,AE}. The shaded band
  displays the isospin-based prediction. The dashed line gives the "modified
  Valencia" calculation \cite{deldel}. The solid line is obtained, if an
  associatedly produced $D_{21}$ resonance is added according to the 
  process $pp \to D_{21}\pi^- \to pp\pi^+\pi^-$ with a strength fitted to the
  total cross section data. 
}
\label{fig4}
\end{figure}

In order to determine the energy dependence of total and differential cross
sections for the quasi-free process, we have divided our background corrected
data into bins of 50 MeV width in the incident energy
$T_p$. The resulting total cross sections are shown in Fig. ~\ref{fig4} (solid
circles) together with results from earlier measurements (open symbols) 
\cite{Brunt,Shimizu,Sarantsev,WB2,Eisner,kek,JJ,JP,AE}. Our
data for the total cross section are in reasonable agreement with the earlier
measurements.

In order to compare with theoretical expectations we plot in Fig.~\ref{fig4}
the results of the "modified Valencia" calculations by the
dashed line. These calculations do very well at low energies, but underpredict
substantially the data at higher energies.
The reason is that by isospin relations $pp\pi^0\pi^0$ and $pp\pi^+\pi^-$
channels have to behave qualitatively similar, if only $t$-channel Roper and
$\Delta\Delta$ processes contribute. So, if the kink around $T_p \approx$ 1.1 
GeV in the $pp\pi^0\pi^0$ data \cite{deldel} got to be reproduced by any such
model calculation, then also the $pp\pi^+\pi^-$ channel has to behave such
(shaded band in Fig. 1), if
not a new strong and very selective $\rho$ channel $\pi^+\pi^-$ production
process enters \cite{D21full}.

\begin{figure} [t]
\begin{center}
\includegraphics[width=0.49\columnwidth]{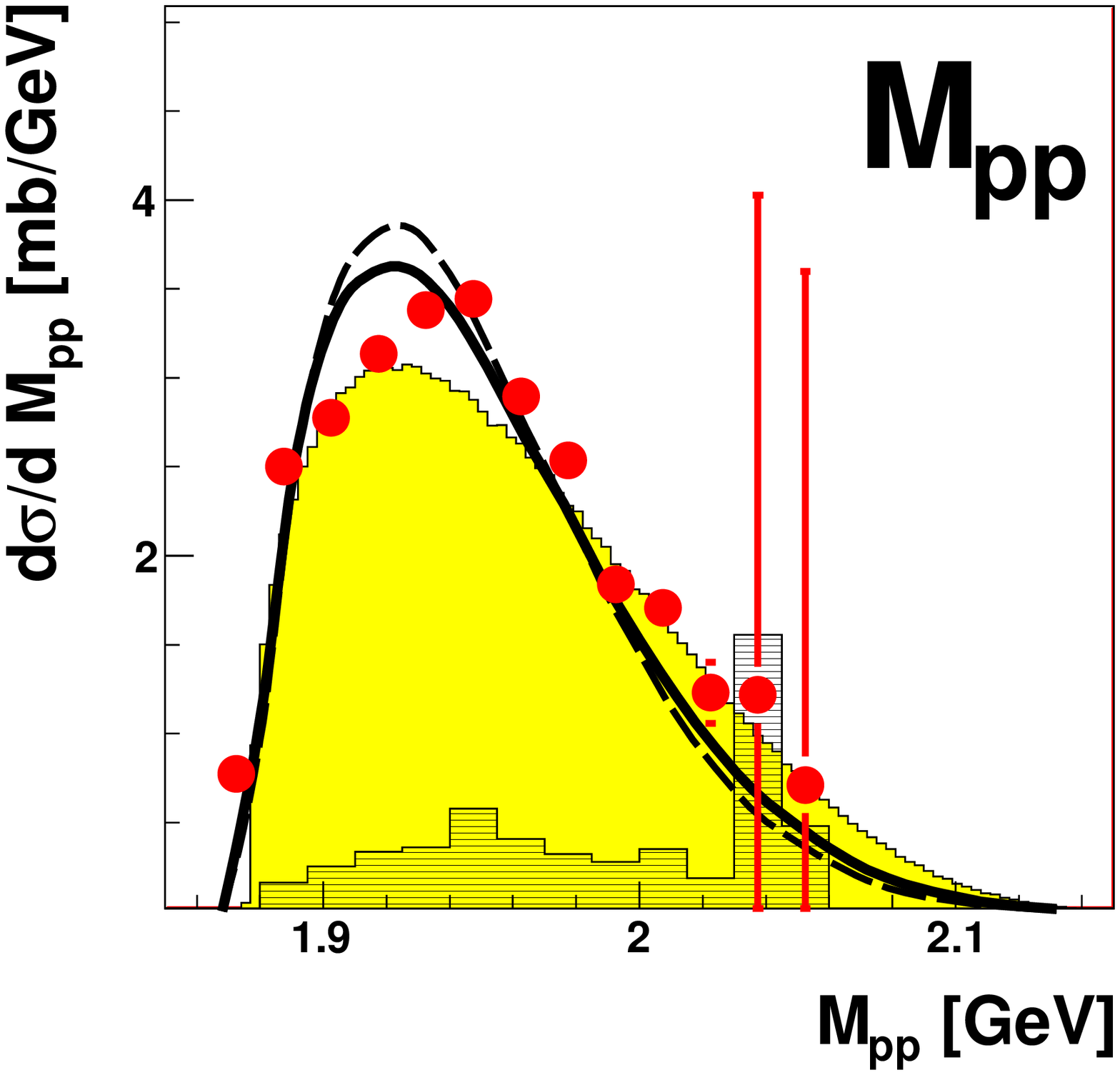}
\includegraphics[width=0.49\columnwidth]{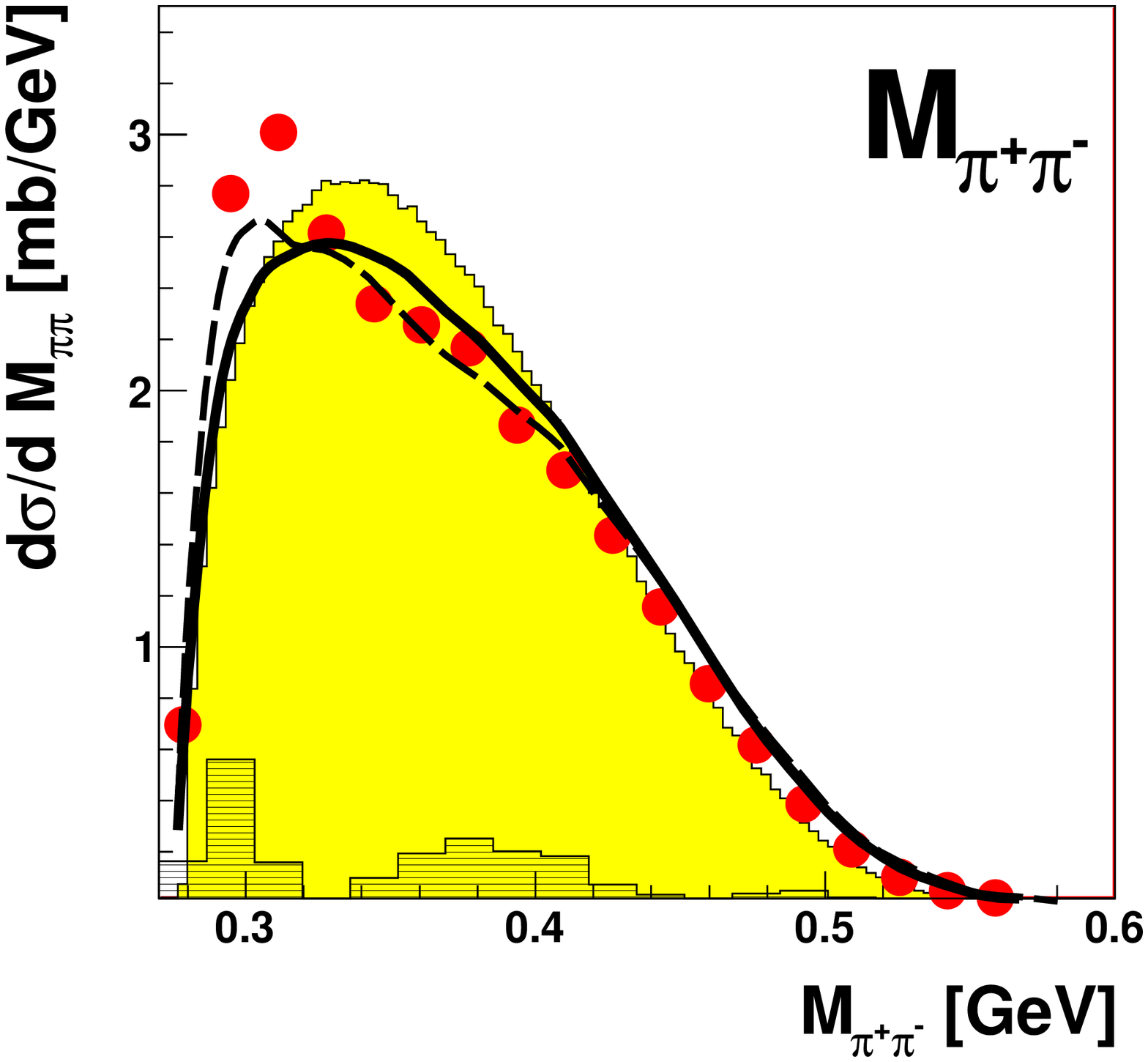}
\includegraphics[width=0.49\columnwidth]{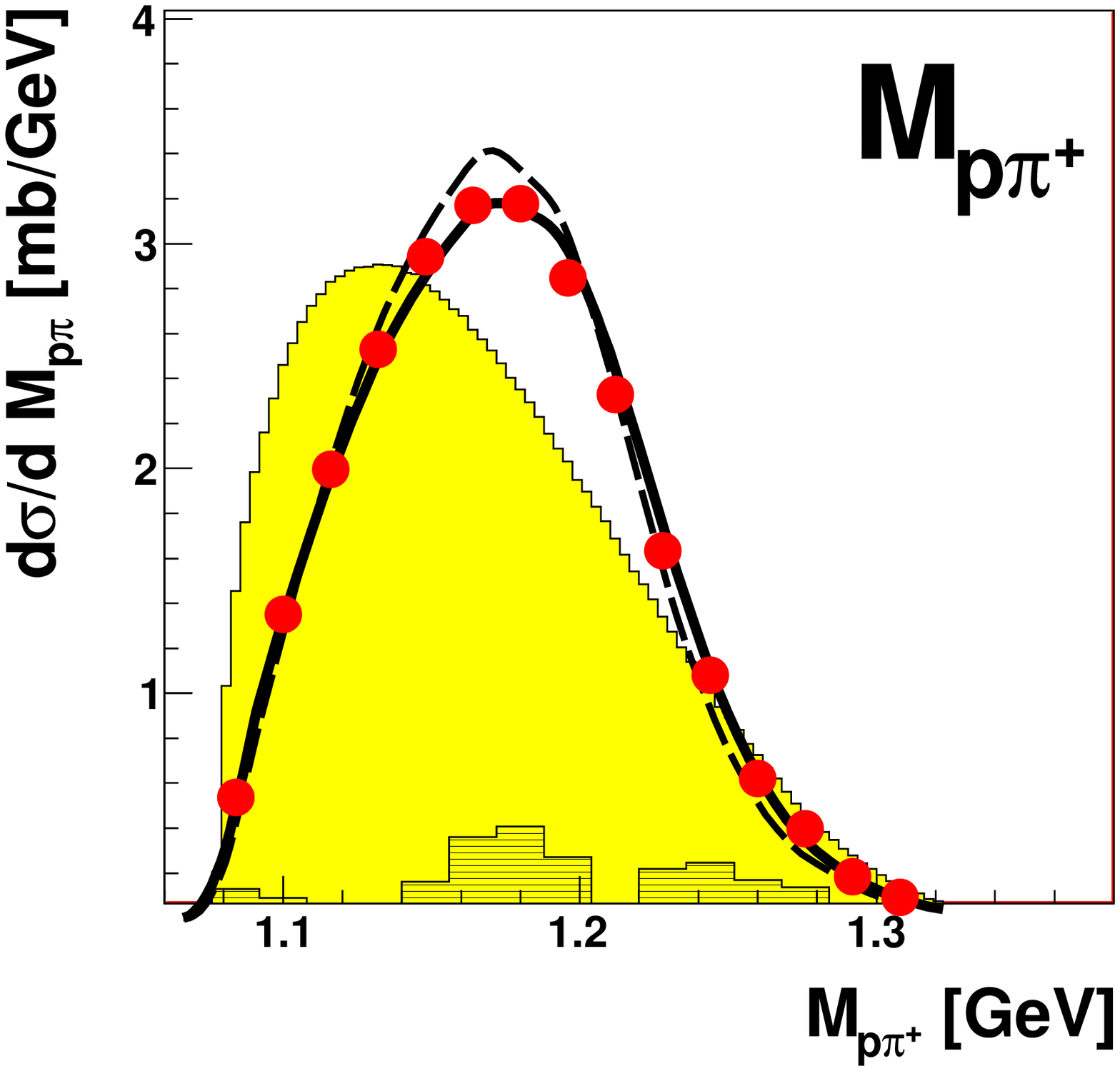}
\includegraphics[width=0.49\columnwidth]{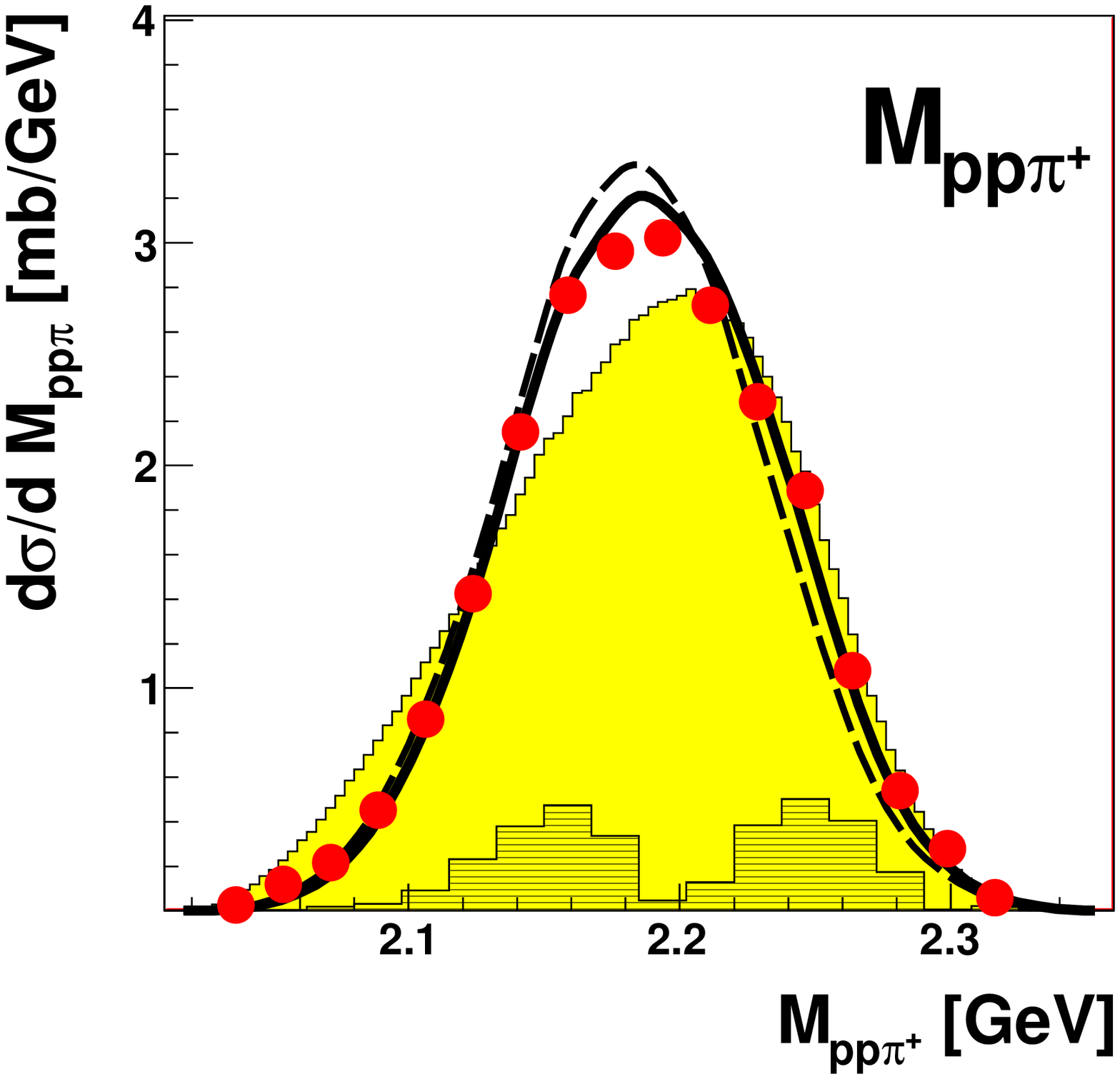}
\includegraphics[width=0.49\columnwidth]{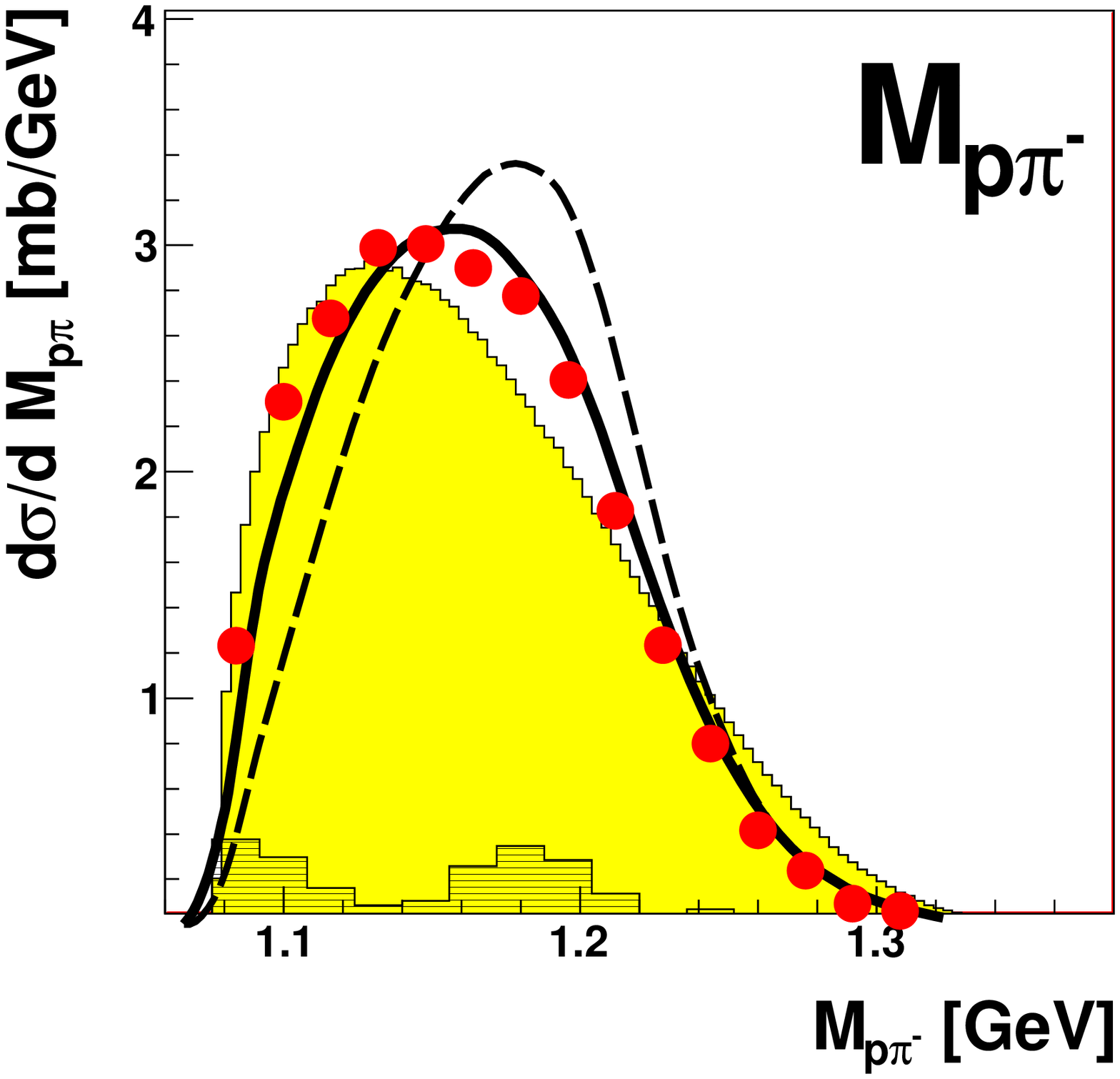}
\includegraphics[width=0.49\columnwidth]{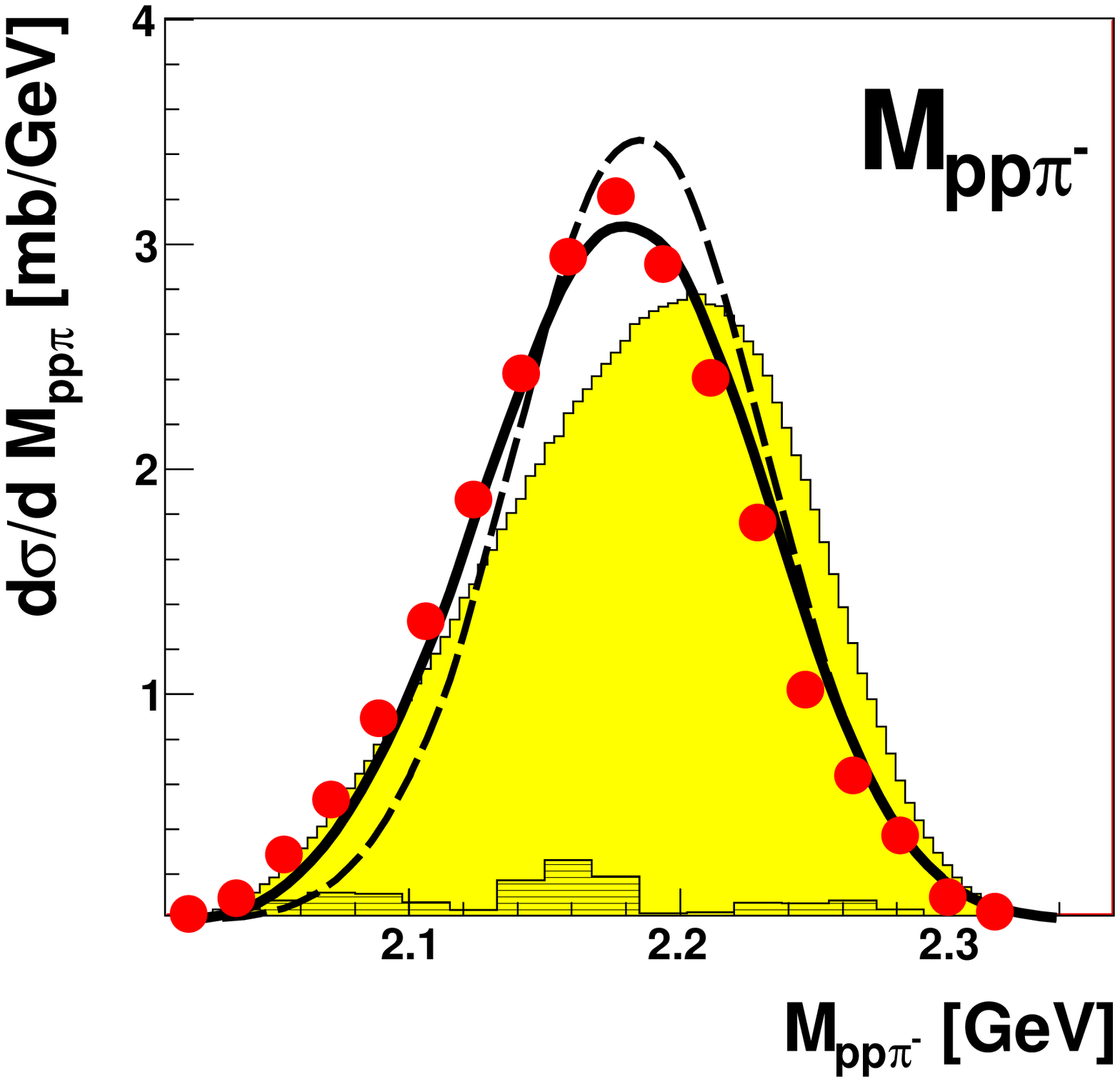}
\caption{(Color online) 
  Differential distributions of the $pp \to pp\pi^+\pi^-$ reaction in the
  region $T_p$ = 0.9 - 1.3 GeV for the invariant-masses $M_{pp}$ (top
  left), $M_{\pi^+\pi^-}$ (top right), $M_{p\pi^+}$ (middle left), $M_{pp\pi^+}$
  (middle right), $M_{p\pi^-}$ (bottom left), $M_{pp\pi^-}$ (bottom right). 
  Filled circles denote the results from this work. The hatched histograms
  indicate 
  systematic uncertainties due to the restricted phase-space coverage of the 
  data. The shaded areas represent pure 
  phase-space distributions, dashed lines 
  "modified Valencia" calculations \cite{Luis} (\cite{deldel}).
  The solid lines include the process $pp \to D_{21}\pi^- \to
  pp\pi^+\pi^-$. All calculations are normalized in area to the data.
}
\label{fig5}
\end{center}
\end{figure}

\begin{figure} 
\begin{center}
\includegraphics[width=0.49\columnwidth]{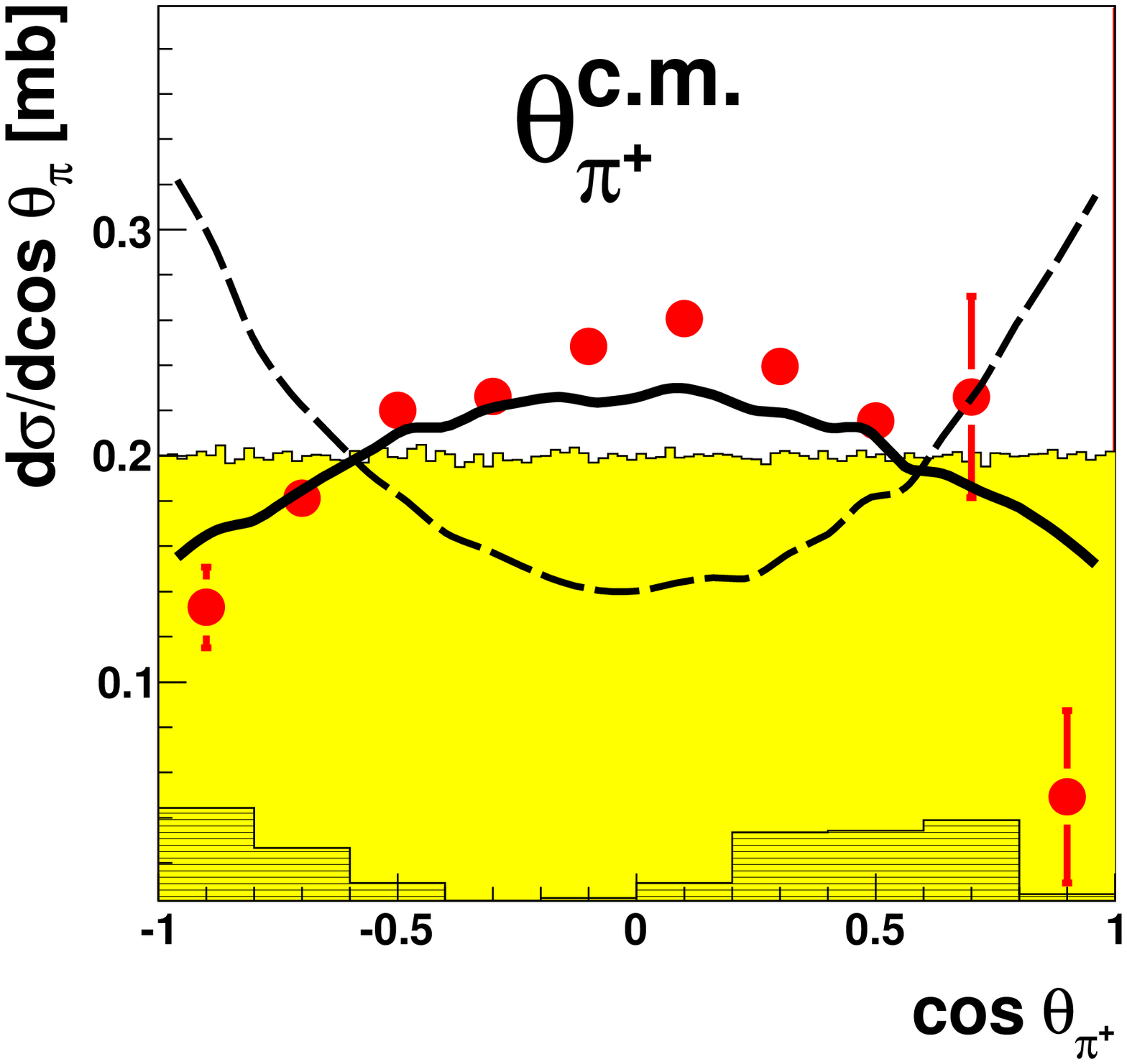}
\includegraphics[width=0.49\columnwidth]{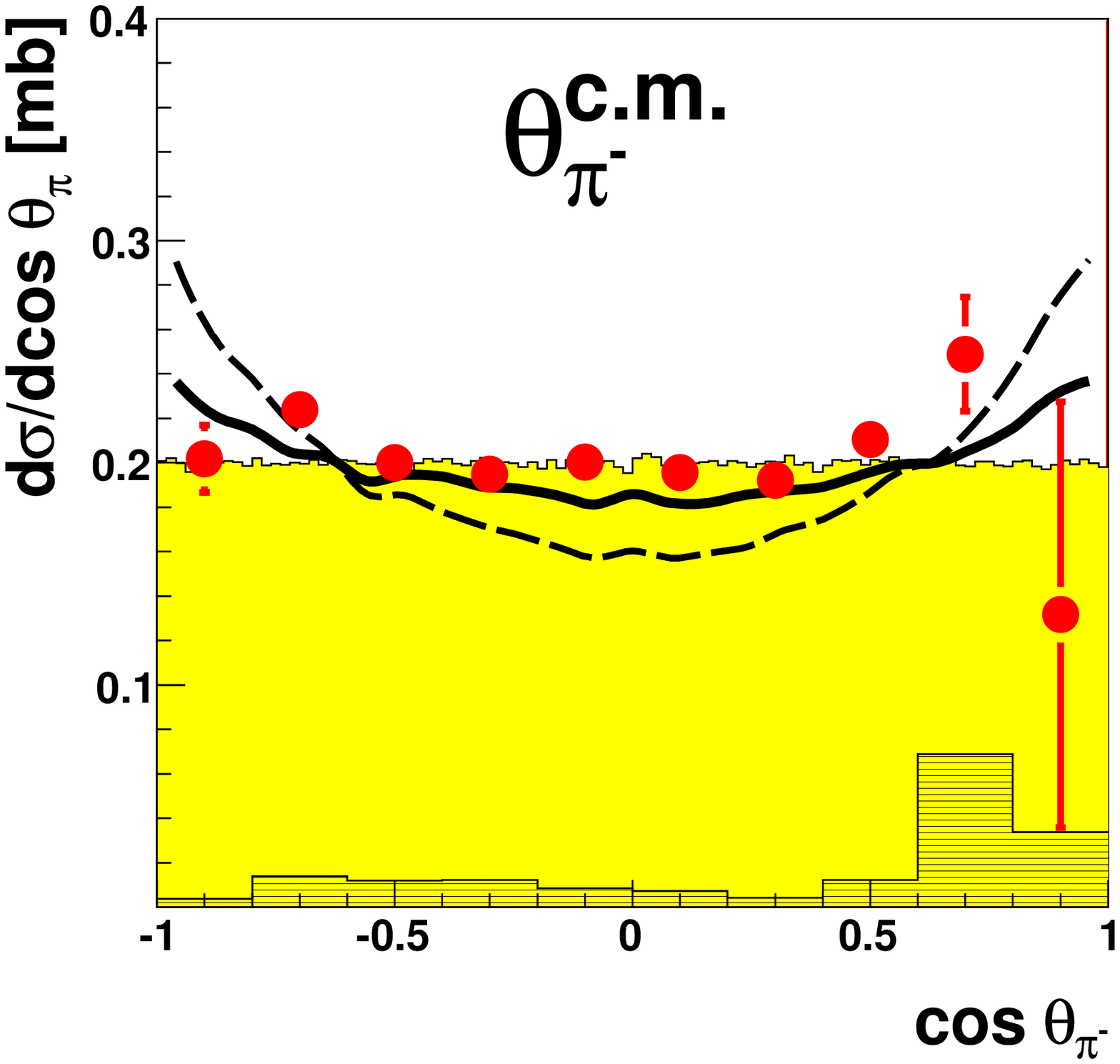}
\includegraphics[width=0.49\columnwidth]{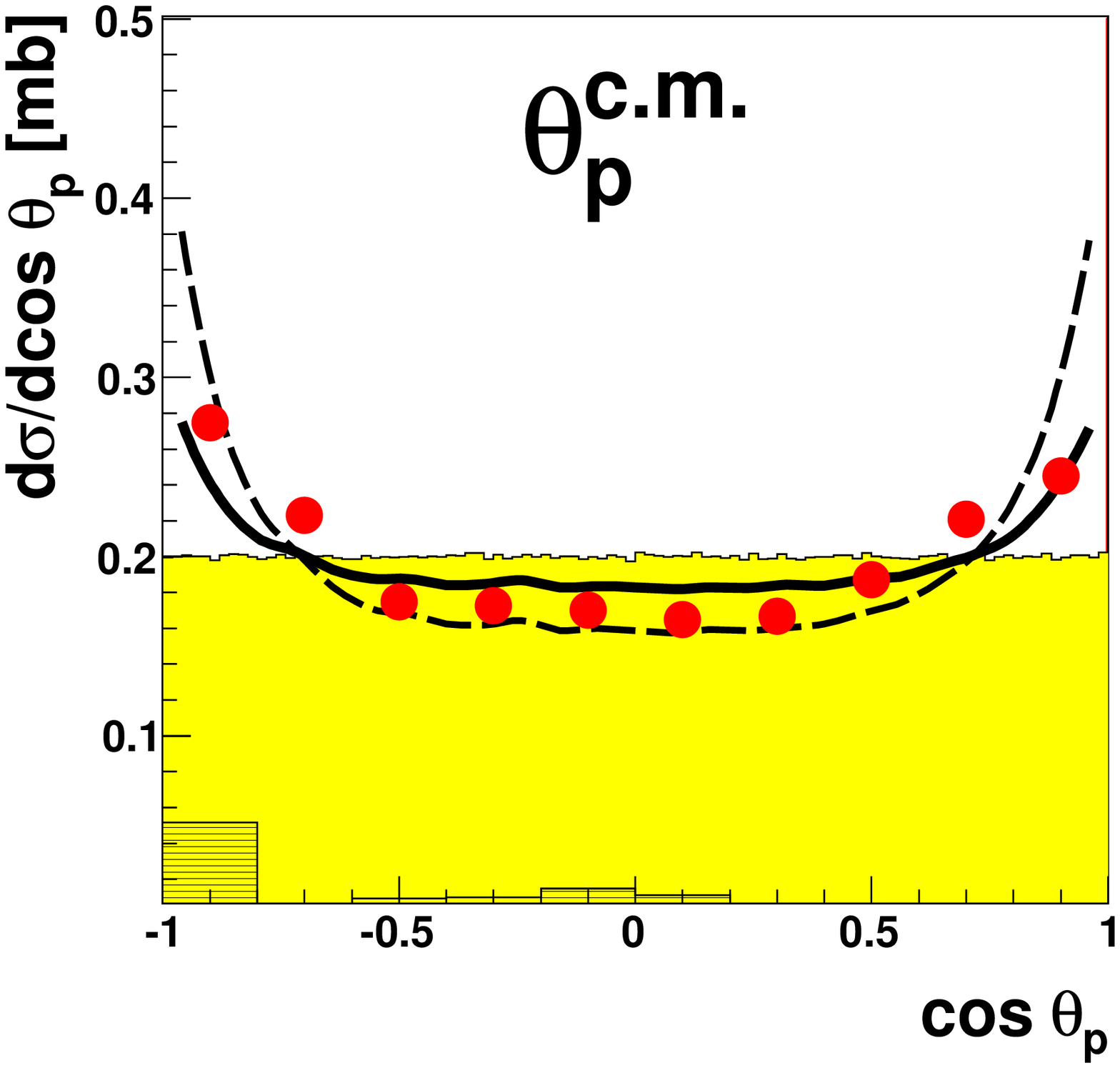}
\caption{(Color online) 
   The same as Fig. ~\ref{fig5}, but for the c.m. angles of  positive and
   negative pions $\Theta_{\pi^+}^{c.m.}$ and $\Theta_{\pi^-}^{c.m.}$,
   respectively, as well as protons $\Theta_p^{c.m.}$.
}
\label{fig6}
\end{center}
\end{figure}


Next we consider the differential cross sections.
For a four-body, axially symmetric final state there are seven independent
differential observables. For a better discussion of the physics issue we
choose to show in this paper nine differential distributions, namely those 
for the center-of-mass (c.m.) angles for protons and pions denoted by 
$\Theta_p^{c.m.}$, $\Theta_{\pi^+}^{c.m.}$ and $\Theta_{\pi^-}^{c.m.}$,
respectively, as well as  those for the invariant masses $M_{pp}$,
$M_{\pi^+\pi^-}$, $M_{p\pi^+}$, $M_{pp\pi^+}$, $M_{p\pi^-}$ and $M_{pp\pi^-}$. These
distributions are shown in Figs.~\ref{fig5}~-~\ref{fig6}.  

There are no data to compare with from previous experiments in the energy
range considered here. All measured differential distributions are markedly
different from pure phase-space distributions (shaded areas in
Figs.~\ref{fig5} - \ref{fig6}). With the exception of $\Theta_{\pi^+}^{c.m.}$,
$M_{p\pi^-}$ and 
$M_{pp\pi^-}$ spectra, the differential distributions are reasonably well
reproduced by the "modified Valencia model" calculations (dashed curves).  
For better comparison all
calculations are adjusted in area to the data in Figs.~\ref{fig5}~-~\ref{fig6}. 

The proton angular distribution is strongly forward-backward peaked as
expected for a peripheral reaction process. The $\pi^-$ angular distribution is
rather flat, in tendency slightly convex curved, as also observed in the other
$NN\pi\pi$ channels in this energy range.  

But surprisingly, the $\pi^+$ angular distribution exhibits an opposite
curvature, a strikingly concave shape. Such a behavior, which is in
sharp contrast to the theoretical expectations, has been observed so far in
none of the two-pion production channels \cite{D21full}.

Also the  $M_{p\pi^-}$ and $M_{pp\pi^-}$ spectra are markedly different from the
$M_{p\pi^+}$ and $M_{pp\pi^+}$ spectra, respectively. In case of the $t$-channel
$\Delta\Delta$ process, which is usually considered to be the dominating one
at the energies of interest here, $\Delta^{++}$ and $\Delta^0$ get excited 
simultaneously and with equal strength. Hence, the $M_{p\pi^-}$
($M_{pp\pi^-}$) spectrum should be equal to the $M_{p\pi^+}$ ($M_{pp\pi^+}$)
one and the $\pi^+$ angular distribution should equal the $\pi^-$ angular
distribution.

This model-independent observation supported by the failure of the "modified
Valencia" calculation to describe properly 
both the total cross section and the differential distributions suggests that
the $t$-channel $\Delta\Delta$ process is not the leading one here.

It looks that an important piece of reaction dynamics is missing, which
selectively affects the $\pi^+$, $p\pi^-$ and $pp\pi^-$ subsystems in the
$pp\pi^+\pi^-$ channel. Since there is no baryon excitation, which could cure
these problems here, and since the 
discrepancy between data and "modified Valencia" description opens up
scissor-like around $T_p \approx$ 0.9 GeV, it matches the opening of a new
channel, where a $\Delta N$ system is produced associatedly with another
pion. In addition the $\Delta N$ system has to be isotensor, in order to have
the $\Delta$ excitation only in ${p\pi^+}$ system as observed in the
data. Such a state with the desired properties could be the isotensor $D_{21}$ 
state with $I(J^P) = 2(1^+)$ predicted already by Dyson and Xoung \cite{Dyson}
with a mass in the region of its isospin partner $D_{12}$ with $I(J^P) =
1(2^+)$.  The latter has been observed with a mass of about 2144 - 2148 MeV
\cite{Hoshizaki,SAID}, {\it i.e.} with a 
binding energy of a few MeV relative to the nominal $\Delta N$ threshold
and with a width compatible to that of the $\Delta$. For a recent discussion
about the nature of this $D_{12}$ state see, {\it e.g.}, Ref. \cite{hcl}. 

Due to its isospin $I$ = 2 $D_{21}$ cannot be reached directly by the initial
$pp$ collisions, but only be produced associatedly with an additional pion. 
The hypothetical isotensor state $D_{21}$ strongly favors
the purely isotensor channel $pp\pi^+$ in its decay. In addition, $J^P = 1^+$
can be easily reached by adding a $p$-wave pion (from $\Delta$ decay) to a
$pp$ pair in the $^1S_0$ partial wave. Hence -- as already suggested by Dyson
and Xuong \cite{Dyson} -- the favored production process should be $pp \to
D_{21}\pi^- \to pp\pi^+\pi^-$. 


Quantitatively the process can be described by using the formalism outlined in
Refs. \cite {ABC,D21full}
   by adding the $D_{21}$ production on the amplitude level. The $D_{21}$
   resonance can be formed together with 
an associatedly produced pion either in relative $s$ or $p$ wave. In the first
instance the initial $pp$ partial wave is $^3P_1$, in the latter one it is
$^1S_0$ or $^1D_2$. The first case is special, since only this one yields a
$sin \Theta_\pi^{c.m.}$ dependence for the angular distribution of the pion
originating  from the $D_{21}$ decay  --- exactly what is needed for the
description of the data for the $\pi^+$ angular distribution being associated
simultaneously with a flat $\pi^-$ angular distribution.

In fact, if we add such a resonance assuming the process $pp \to D_{21}\pi^-
\to pp\pi^+\pi^-$ with fitted mass $m_{D_{21}}$ = 2140 MeV and width 
$\Gamma_{D_{21}}$ = 110 MeV, we obtain a good
description of the total cross section by adjusting the strength of the
assumed resonance process to the total cross section data (solid line in
Fig.~\ref{fig4}). Simultaneously, the 
addition of this resonance process provides a quantitative description of all 
differential distributions (solid lines in Figs.~~\ref{fig5} - \ref{fig6}), in
particular also 
of the $\Theta_{\pi^+}^{c.m.}$, $M_{p\pi^-}$ and $M_{pp\pi^-}$ distributions. Since
the $D_{21}$ decay populates only $\Delta^{++}$, its reflexion in the
$M_{p\pi^-}$ spectrum shifts the strength to lower masses -- as required by
the data. The same holds for the $M_{pp\pi^-}$ spectrum. 

We note that the only other place in pion production, where a
concave curved pion angular distribution has been observed, is the $pp \to
pp\pi^0$ reaction in the region of single $\Delta$ excitation
\cite{ED,ANKE}. Also in this case it turned out that the reason for
it was the excitation of resonances in the $\Delta N$ system \cite{ANKE}
causing a proton spinflip situation as in our case here.

Though the addition of an isotensor dibaryon resonance cures the
shortcomings of the "modified Valencia" calculations for the $pp \to
pp\pi^+\pi^-$ reaction, we 
have to investigate, whether such an addition leads to inconsistencies in the
description of other two-pion production channels, since 
such a state may decay
also into $NN\pi$ channels other than $pp\pi^+$  --- though with a much
smaller branchings due to isospin coupling. In consequence it may also
contribute to other two-pion production channels. This is particularly relevant
for the $pp \to pp\pi^0\pi^0$ reaction with its comparatively small cross
section at the energies of interest here. But the $D_{21}$ production via the
$^3P_1$ partial wave leaves the two pions in relative $p$-wave, hence they are
also in an isovector state by Bose symmetry. Since such a $\rho$-channel
situation is not possible for identical pions, there are no contributions from
$D_{21}$ in $pp\pi^0\pi^0$ and $nn\pi^+\pi^+$ channels, {\it i.e.} there is no
consistency problem.

From a fit to the data we obtain a mass $m_{D_{21}}$ = 2140(10) MeV and a width
$\Gamma_{D_{21}}$ = 110(10) MeV. The mass is in good agreement with the
prediction of Dyson and Xuong \cite{Dyson}. Both mass and width are just
slightly smaller than those calculated by Gal and Garcilazo \cite{GG}.

\section{Summary and Conclusions}

Total and differential cross sections of the $pp \to pp\pi^+\pi^-$ reaction
have been measured exclusively and kinematically complete in the energy range
$T_p = 1.08 - 1.36$ GeV 
by use of the quasi-free
process $pd \to pp\pi^+\pi^- + n_{spectator}$. The results for the total
cross section are in good agreement with previous bubble-chamber data. For the
differential cross sections no data from previous measurements are
available.

The $M_{p\pi^-}$, $M_{pp\pi^-}$ and $\Theta_{\pi^-}^{c.m.}$ distributions are
observed to be strikingly different from their counterparts, the $M_{p\pi^+}$,
$M_{pp\pi^+}$ and $\Theta_{\pi^+}^{c.m.}$ distributions, respectively. Hence
the originally anticipated $t$-channel $\Delta\Delta$ mechanism cannot be the
dominating process here.
 
The problem can be overcome, if there is an opening of a new reaction
channel near $T_p \approx$ 0.9 GeV, {\it i.e.}, near the $\Delta N\pi$
threshold, which nearly exclusively feeds the
$pp\pi^+\pi^-$ channel. Such a process is the associated production of
the theoretically predicted isotensor $\Delta N$ state $D_{21}$ with specific
signatures in invariant 
mass spectra and in the $\pi^+$ angular distribution. We have
demonstrated that such a process provides a quantitative description of the data
for the $pp \to pp\pi^+\pi^-$ reaction --- both for the total cross section
and for all differential distributions.


\section{Acknowledgments}

We acknowledge valuable discussions with A. Gal, Ch. Hanhart, V. Kukulin and
G. J. Wagner on this 
issue. We are particularly indebted to L. Alvarez-Ruso for using his code. 
This work has been supported by DFG (CL 214/3-2) and STFC (ST/L00478X/1) as
well as by the Polish National Science Centre through the grant
2016/23/B/ST2/00784.

\end{document}